\documentclass[12pt,preprint]{aastex}
\usepackage[dvipdfm, hyperindex=true, colorlinks=true, linkcolor=blue, citecolor=blue, anchorcolor=blue]{hyperref}

\newcommand{\SM}{\,{\ensuremath{M_{\odot}}}}
\newcommand{\SMPY}{\,{\ensuremath{M_{\odot}\,{\rm yr^{-1}}}}}
\newcommand{\es}{\,{\ensuremath{\rm erg\,{s}^{-1}}}}

\def\DP#1{{\ensuremath{10^{\rm #1}}}}
\def\TDP#1{{\ensuremath{\times 10^{\rm #1}}}}

\begin{document}

\title{On the evolution of a fossil disk around neutron stars originating from merging WDs}
\author{Bai Sheng Liu$^{1,2}$ and Xiang-Dong Li$^{1,2}$}

\affil{$^{1}$Department of Astronomy, Nanjing University, Nanjing
210046, China}

\affil{$^{2}$Key laboratory of Modern Astronomy and Astrophysics
(Nanjing University), Ministry of Education, Nanjing 210046, China}

\affil{$^{}$lixd@nju.edu.cn}

\begin{abstract}

Numerical simulations suggest that merging double white dwarfs (WDs)
may produce a newborn neutron star surrounded by a fossil disk. We
investigate the evolution of the fossil disk following the
coalescence of double WDs. We demonstrate that the evolution can be
mainly divided into four phases: the slim disk phase (with time
$\lesssim$ 1 yr), the inner slim plus outer thin disk phase
($\sim 10-\DP{6}$ yr), the thin disk phase ($\sim \DP{2}-\DP{7}$
yr), and the inner advection-dominated accretion flow plus outer
thin disk phase, given the initial disk mass $\sim 0.05-0.5\,M_{\sun}$ and the disk formation time $10^{-3}-1$ s. Considering possible wind mass loss from the disk,
we present both analytic formulae and numerically calculated results for
the disk evolution, which is sensitive to the condition that
determines the location of the outer disk radius.  The systems are
shown to be very bright in X-rays in the early phase, but quickly
become transient within $\lesssim$ 100 yr, with peak luminosities
decreasing with time. We suggest that they might account for part of
the very faint X-ray transients around the Galactic center region,
which generally require a very low mass transfer rate.

\end{abstract}

\keywords{accretion, accretion disks$-$stars: evolution$-$stars: neutron$-$white dwarfs$-$X-rays: binaries}

\section{Introduction}
\label{sec.1}

White dwarfs (WDs) are the most common stellar remnants in the
universe. Within 20 pc of the Solar system, about 25\% of the WD
population are in binary systems, and $\sim$ 6\% of them are double
WDs \citep{HolbergSO08}. WDs in compact binaries are responsible for
lots of interesting astrophysical phenomena. For example, due to the
gravitational radiation, the double degenerate stars will gradually
lose their orbital angular momenta and undergo a merging process at
last. Thus, the coalescence of double WDs is a good source of strong
gravitational wave radiation
\citep[e.g.,][]{FryerHH02,Ott09,YuJef10,KilicBG14}.

The merger remnants of double WDs can be various kinds of exotic
objects, depending on the merging process and the initial conditions
of the WDs including the total WD masses
\citep{vanKerkCJ10,vanKerk13}, the binary mass ratio, the
synchronization timescale \citep[e.g.,][]{ZhuCvan13}, and the mass
loss during the merging events
\citep[e.g.,][]{DessartBOL06,DanRoss11,RaskinK13}. If the remnant
mass is less than the Chandrasekhar mass limit ($M_{\rm Ch}$), the
outcome may be a sdB/sdO star
\citep[e.g.,][]{SaioJ00,HanPod02,Heber09}, or a R Coronae Borealis
star \citep[e.g.,][]{Webbink84,JefferyKS11,LonglandLJ11,ZhangJ14}.
This merger scenario has been proposed to explain the bright,
massive ($\gtrsim 0.8{\SM}$) WDs in the Galactic halo
\citep[e.g.,][]{Yuan92,LiebertBH05}, the existence of a disk around
a WD with a metal-rich atmosphere \citep[e.g.,][]{GarciaL07}, and
WDs with a strong (around $10^{6-9}$ G) magnetic field
\citep[e.g.,][]{Garcia12,KulebiE13}.

If the merger remnant is more massive than $M_{\rm Ch}$,  the system
is expected to either explode as a Type Ia supernova \citep[SN
Ia,][]{Webbink84,IbenT84}, or collapse to a neutron star (NS)
\citep[i.e., accretion-induced collapse or
AIC,][]{SaioN85,RuedaBI13}\footnote{Note that besides the above
double degenerate model, the single degenerate one
\citep{WhelanI73,NomotoK91} is another route to produce the above
events.}. During the merger of double WDs, both mass and angular
momentum are accreted onto the more massive WD. In the AIC case, the
newborn NS, may be rapidly rotating and behave as a normal pulsar
(PSR) or a millisecond pulsar \citep[MSP; e.g.,][]{BhattaHeu91}. Moreover, if the magnetic field of the progenitor is strong
enough, it may be amplified via dynamo action and flux conservation
during the collapse \citep[][]{DuncanT92}, and hence the end product
could be a magnetar \citep[e.g.,][]{Usov92,KingPW01,LevanWC06}.

Both smoothed particle hydrodynamics
\citep{BenzBCP90,YoonPod07,LorenIsern09} and grid-based techniques
\citep{SouzaMTF06,MotlFT07} have been adopted to investigate the
coalescence processes of double WDs. A delayed explosion is
generally observed. The less massive WD is disrupted, and a hot,
rotating corona, surrounded by a thick disk, forms around the more
massive WD. The outcome of the CO WD merger depends on whether the
C-ignition occurs at the center \citep{SaioN85,SaioN98,SaioN04} or
in the envelope of the accreting WD \citep{NomotoI85}, which may be
determined by the accretion rate \citep{NomotoI85,KawaiSN87}. If the
ignition occurs at the center, a SN Ia will follow, provided that
the accretion rate is less than {2.7\TDP{-6}\SMPY}. If the C-burning
starts from the outer layer, the object will collapse to be a NS.

According to numerical simulations of WD mergers leading to AIC
\citep{DessartBOL06,DessartBLO07}, a fossil disk of mass $\sim
0.05-0.5${\SM} is left around the newborn NS. \citet{MetzgerPQ09b}
calculated the evolution of the composition of the accretion disk,
and found that no other intermediate-mass elements are synthesized
unlike SN Ia events, and strong winds blow away some disk mass
containing $\sim \DP{-2}\SM$ $^{\rm 56}$Ni, which may explain some
sub-luminous SN Ia events. \citet{DarbhaMQ10} showed that the light
curves and spectra may help distinguish the WD merging events from
other transient events, e.g., .Ia and binary NS mergers.

In this work we focus on the evolution of the fossil disks around
the NSs following the WD coalescence, which is similar to that of
supernova fallback disks \citep[e.g.,][]{MenouPH01,ShenM12}. While
most of the previous studies pay attention to the disk evolution at
the early, bright stage, our work extends the evolution to the late,
faint phase. In addition, we consider the influence on the disk
evolution of both neutrality in the disk and self-gravity of the disk, which are seldom noticed in previous studies.

This paper is organized as follows. We model the evolution of the
fossil disk around a NS following an AIC event in Section 2.
Section 3 presents numerically calculated results based on the model
parameters. We discuss the possible observational implications of
the fossil disk model in Section 4 and summarize in Section 5.

\section{The Model}
\label{sec.2}

Although the evolution of the fossil disks in the AIC model is still under discussion, the studies on the evolution of the fallback disks originating from tidal disruption
events (TDEs) and SNe could be good references, as illustrated by their similar formation processes. During a TDE, as a star is tidally disrupted by a supermassive black hole, roughly half of its debris falls back and orbits the black hole, gradually forms a disk  and accretes onto the hole \citep[e.g.,][]{Rees88}. When a NS is produced in the SN explosion, the progenitor's envelope is ejected outwards; part of the ejecta may be gravitationally captured by the NS and form a fallback disk around it if there is sufficient angular momentum \citep[e.g.,][]{Michel88,Chevalier89}. In the AIC scenario, during the coalescence of double WDs, a fossil disk or torus, which originates from the disrupted, less massive WD, is also observed to be left over around the newborn NS \citep[e.g.,][]{SaioN04,DessartBOL06}. Therefore the evolution of these fossil disks is likely to follow similar rules.
The whole
evolution can be divided into the following four phases. (i) In
phase 1, the whole, newly formed disk is geometrically slim or even
thick with a super-Eddington accretion rate, where advection
dominates cooling \citep[e.g.,][]{CannizzoGeh09}. (ii) In phase 2,
along with the decrease of the accretion rate and the expansion of
the disk, the outer part of the disk gradually turns to be
geometrically thin, where radiative cooling becomes effective, that
is, the disk contains an inner slim region and an outer thin region,
with the latter advancing inward. (iii) In phase 3, the whole disk
becomes optically thick and geometrically thin. (iv) In phase 4, the
inner region of the disk begins to become advection-dominated
accretion flow (ADAF) where radiative cooling is inefficient, which
is surrounded by a outer thin disk. The ADAF region develops
outwards with decreasing accretion rate.

In the following we discuss the disk evolution in detail.

\subsection{Initial Parameters}
\label{sec.2.1}

We consider a NS with mass $M=1.4{\SM}$ and radius ${R}_{\rm NS}=\DP{6}$\,cm. The accretion rate
$\dot{M}$ is expressed in units of the Eddington accretion rate
$\dot{M}_{\rm Edd}$, i.e., $\dot{m}=\dot{M}/\dot{M}_{\rm Edd}$. Here
$\dot{M}_{\rm Edd}=10L_{\rm Edd}/c^2=1.39\TDP{18}m$ gs$^{-1}$ with
$m=M/{\SM}$ and $L_{\rm Edd}=1.25{\TDP{38}m\es}$. The interesting
radius $R$ in the disk is scaled with the Schwarzschild radius
${R}_{\rm S}=2GM/c^2=2.95\TDP{5}m$ cm, i.e., $r=R/R_{\rm S}$. The
formation time $t_{\rm f}$ of the disk ranges from $\sim$ 1.0 ms to
1.0 s, and the initial disk mass $M_{\rm D,0}=\eta\SM$ with
$\eta=0.05-0.5$, depending on the total mass, the initial mass ratio
$q$, and the mass loss during AIC
\citep[e.g.,][]{DessartBOL06,DanRoss11,RaskinK13,ZhuCvan13}, etc.
Generally smaller $q$ (if $q < 1$) may lead to a larger $M_{\rm
D,0}$ \citep{MetzgerPQ08}. The mass distribution of the disk is very
close to that of NS + NS or NS + black hole (BH) mergers
\citep[e.g.,][]{MetzgerPQ09a}, so the model may also be applicable
to those cases.

The inner radius $r_{\rm i}$ and the outer radius $r_{\rm out}$ of
the disk at $t_{\rm f}$ are assigned to be $r_{\rm i} \gtrsim$ 2.5
(roughly the radius of the newborn NS) and $r_{\rm out}=r_{\rm f}
\sim 10^{3-4}$ (roughly the radius of the accreting WD),
respectively. We assume that the NS magnetic field is very weak
($\lesssim \DP{8-9}$ G), so the disk can reach the surface of the
NS. In the future work we will consider the interaction of a
strong magnetic field with the disk. The outer disk spreads due to
angular momentum transport \citep{LyndenP74,Pringle91}, until it reaches a radius where self-gravity becomes important. The stability
criterion for a differentially rotating disk is
$Q\equiv\frac{c_{\rm s}\Omega_{\rm K}}{\pi G\Sigma}=1$,
where $c_{\rm s}$, $\Omega_{\rm K}$, and $\Sigma$ are the sound speed, the Keplerian angular velocity, and the surface density of the disk material, respectively, and $G$ is the gravitational constant \citep{Toomre64}. This defines
the self-gravity radius as \citep{BurderiKS98},
\begin{eqnarray}
  r_{\rm sg} \simeq  2.92\TDP{5}\cdot{m}^{-2/3}\cdot(\frac{\rho}{1\,{\rm gcm}^{-3}})^{-1/3}, \label{eqn.1}
\end{eqnarray}
where $\rho$ is the density of the disk material. Note that the magnitude of the self-gravity radius is very similar to the Roche limit \citep[e.g.,][]{AggOb74,MenouPH01}, and we will see that it is a crucial factor in the disk evolution.

As the accretion rate decays, the energy dissipation rate in the
disk decreases. The outermost disk, which is the coldest, will become
neutral eventually once the temperature is lower than a critical
value $T_{\rm P}\sim 6500$ K \citep[][]{MenouPH01,ErtanEE09}. After
that there is hardly any transport of mass or angular momentum in
the disk. However, it has been suggested the irradiation from the central
source may prevent the appearance of a passive disk even at a lower
temperature \citep[][]{Alpar01,InotsukaS05,AlparCE13}. The disk
evolution, after being neutral, is less certain and beyond our
interest. In the following calculations, the outer disk is assumed
to be neutral and unchanging if its temperature drops to $T_{\rm
P}=300$ K \citep[][]{Alpar01}.

\subsection{Disk Evolution}
\label{sec.2.2}

After the coalescence of double WDs, a fossil disk originating from
the ejecta is formed around the NS with a timescale $\sim$ 1.0
ms$-$1.0 s \citep{DessartBOL06}. Along with mass accretion, the disk
spreads to a larger radius due to the conservation of angular
momentum.
The evolution of the surface density $\Sigma$ of the fossil disk is governed by a linear, viscous diffusion equation \citep[][]{Pringle81,CannizzoLee90},
\begin{eqnarray}
\frac{\partial\Sigma}{\partial{t}}=\frac{3}{R}\frac{\partial}{\partial{R}} \left[R^{1/2}\frac{\partial}{\partial{R}}(\nu\Sigma{R^{1/2}})\right],\label{eqn.2}
\end{eqnarray}
where $\nu=2\alpha{P}/(3\rho\Omega_{\rm K})$ is the  kinetic
viscosity coefficient. Here $P$ is the pressure and $\alpha$ is the standard
\citet{ShakuraS73} viscosity parameter. If $\nu$ follows a power law
of $R$, $\nu \propto {R^{n}}$, the above equation can be solved
analytically with self-similarity solutions
\citep[][]{Pringle74,Pringle81}. The evolution of the disk can be
classified into the following phases.

In phase 1, the accretion rate is usually so high that the disk is described by a slim disk model, where radiation pressure dominates and advection can effectively consume most of the heat locally generated  \citep{AbramowiczCLS88}.
While the fallback rate evolves with time $\propto {t}^{-5/3}$ \citep[][]{Phinney89}, the accretion rate linked to viscous diffusion follows $\propto {t}^{-4/3}$.
Meanwhile, in the slim disk phase possible disk wind loss should be taken into account, so the local accretion rate is assumed to follow a power law of the radius, $\dot{m}(r) \propto r^{s}$, where the power-law index $s$ ($0 < s < 1$) is believed to be constant \citep[][]{BlandBegel99,NarayanPK01}.
If the wind loss is weak ($s<1/3$), the accretion rate can be described by $\propto {t}^{-4/3-s}$, otherwise $\propto {t}^{-5/3}$ \citep{Pringle91,CannizzoGeh09,ShenM12}. Thus the parameters of the disk vary as,
\begin{eqnarray}
   && \dot{m}(r,t) = \left\{ \begin{array}{ll}
  \dot{m}_{0}(\frac{r}{r_{\rm f}})^{s}(\frac{t}{t_{\rm f}})^{-4/3-s}  & (0<s<1/3),\\
  \dot{m}_{0}(\frac{r}{r_{\rm f}})^{s}(\frac{t}{t_{\rm f}})^{-5/3}    & (1/3<s<1),
  \end{array}
  \right.\nonumber\\
  && r_{\rm out}(t) = r_{\rm f}(\frac{t}{t_{\rm f}})^{2/3},\nonumber\\
  && T(r,t)=T_{0}(\frac{\dot{m}}{\dot{m}_{0}})^{1/4}{(\frac{r}{r_{\rm f}})^{-5/8}}, \nonumber\\
  && \rho(r,t)=\rho_{0}(\frac{\dot{m}}{\dot{m}_{0}}){(\frac{r}{r_{\rm f}})^{-3/2}}, \label{eqn.3}
\end{eqnarray}
where $T$ is the temperature,  and the subscript 0 denotes variables measured at $t=t_{\rm f}$ and $r=r_{\rm f}$. From \citet{CannizzoGeh09}
we have
\begin{eqnarray}
  && T_{0}\simeq 1.0\TDP{8}({\alpha m\over0.1})^{-1/4} \dot{m}_{0}^{1/4}\cdot {r}_{\rm f}^{-5/8}\,\,{\rm K},\nonumber\\
  && \rho_{0}\simeq 3.0\TDP{-4}({\alpha m\over0.1})^{-1} \dot{m}_{0}\cdot{r_{\rm f}^{-3/2}}\,\,{\rm gcm}^{-3}, \label{eqn.4}
\end{eqnarray}
where $\dot{m}_{0}$ can be solved from the initial
mass of the disk,
\begin{eqnarray}
  && M_{\rm D,0}=\eta\SM=\dot{m}_{0}\dot{M}_{\rm Edd}{t}_{\rm f}.\label{eqn.5}
\end{eqnarray}

Phase 1 ends at the time $t_{1}$  when radiative cooling begins to dominate in the outermost region, that is, the transition radius $r_{\rm tra}$, which separates the inner slim disk region from the outer radiative cooling-dominated region, starts to equal the outer disk radius. The slim disk region is characterized by both advective cooling and radiation pressure domination. The boundary condition for the slim-thin disk transition occurs when the ratio of the advective cooling and the energy generation via  viscous dissipation becomes less than $1/3-1/2$ \citep[e.g.,][]{AbramowiczCLS88,Kato98}. Detailed calculations show that advection starts to modify the disk structure at local accretion rates $\sim 0.6-0.8\dot{M}_{\rm Edd}$ \citep{WataraiFT00,Sadowski11}. So we can roughly estimate $r_{\rm tra}$ as
\begin{equation}
   r_{\rm tra} \simeq \dot{m}(r_{\rm out}, t)=r_{\rm out}.\label{eqn.6}
\end{equation}
This yields
\begin{eqnarray}
  && t_1= \left\{ \begin{array}{ll}
  t_{\rm f}\cdot(\frac{\dot{m}_{0}}{r_{\rm f}})^{3/(6+s)}  & (0<s<1/3),\\
  t_{\rm f}\cdot(\frac{\dot{m}_{0}}{r_{\rm f}})^{3/(7-2s)} & (1/3<s<1).
  \end{array}
  \right.\label{eqn.7}
\end{eqnarray}

In phase 2, The thin disk region develops inwards till the inner slim disk region disappears. While the inner slim disk region is always dominated by radiation pressure and electron scattering opacity, the outer thin disk region will undergo two status-transitions with decreasing accretion rate. Firstly, the outer, thin disk region, initially dominated by radiation pressure, gradually becomes supported by gas pressure. Secondly, the electron scattering opacity in the thin disk region is replaced by free-free absorption. The evolution
can be separated by the time point $t_{\rm gas}$ when
\begin{eqnarray}
  && r_{\rm gas} \simeq 281\cdot({\alpha m\over 0.1})^{2/21}\dot{m}^{16/21}=r_{\rm out}, \label{eqn.8}
\end{eqnarray}
where the radius $r_{\rm gas}$ separates the region dominated by gas
pressure from that dominated by radiation pressure \citep{ShakuraS73}.

Before the first status-transition ($t_{1} < t < t_{\rm gas}$), the whole disk comprises an inner, slim disk region and an outer, thin disk region (the entire disk is dominated by radiation pressure plus electron scattering opacity). Taking into account the fact that $\dot{m}(r) \propto r^{s}$ in the slim disk region and $\dot{m}(r) \propto t^{-8/7}$ in the thin disk region \citep{ShenM12} and the boundary condition between them, one can obtain the following formulae for the disk evolution,
\begin{eqnarray}
   && \dot{m}(r,t) = \left\{ \begin{array}{ll}
  k_{1}(\frac{t}{t_{\rm f}})^{-8(1-s)/7} (\frac{r}{r_{\rm f}})^{s} & \mbox{if $r_{\rm i} < r < r_{\rm tra}$},\\
  k_{2} (\frac{t}{t_{\rm f}})^{-8/7}      & \mbox{if $r_{\rm tra} < r < r_{\rm out}$},
  \end{array}
  \right.\nonumber\\
  && r_{\rm out}(t) = k_{3}(\frac{t}{t_{\rm f}})^{2/7},\nonumber\\
  && T(r,t) = \left\{ \begin{array}{ll}
  T_{1}(\frac{\dot{m}}{\dot{m}_{0}})^{1/4}{(\frac{r}{r_{\rm f}})^{-5/8}} & \mbox{if $r_{\rm i} < r < r_{\rm tra}$},\\
  T_{2}(\frac{r}{r_{\rm f}})^{-3/4}      & \mbox{if $r_{\rm tra} < r < r_{\rm out}$},
  \end{array}
  \right.\nonumber\\
  && \rho(r,t) = \left\{ \begin{array}{ll}
  \rho_{1}(\frac{\dot{m}}{\dot{m}_{0}})(\frac{r}{r_{\rm f}})^{-3/2}      & \mbox{if $r_{\rm i} < r < r_{\rm tra}$},\\
  \rho_{2}(\frac{\dot{m}}{\dot{m}_{0}})^{-2}(\frac{r}{r_{\rm f}})^{3/2} & \mbox{if $r_{\rm tra} < r < r_{\rm out}$},
  \end{array}
  \right.\label{eqn.9}
\end{eqnarray}
where
\begin{eqnarray}
  && k_{1}=r_{\rm f}(\frac{t_1}{t_{\rm f}})^{38(1-s)/21}, \quad k_{2}=r_{\rm f}(\frac{t_1}{t_{\rm f}})^{38/21}, \quad k_{3}=r_{\rm f}(\frac{t_1}{t_{\rm f}})^{8/21},\nonumber\\
  && T_{1}=T_{0}, \quad T_{2}=T_{0} (\frac{t_{\rm 1}}{t_{\rm f}})^{1/4} (\frac{r_{\rm f}}{\dot{m}_{0}})^{1/4}, \quad {\rho}_{1}={\rho}_{0}, \quad {\rho}_{2}={\rho}_{0}(\frac{r_{\rm f}}{\dot{m}_{0}})^{3}, \nonumber\\
  && t_{\rm gas}=t_{\rm f} [281\cdot{({\alpha m\over 0.1})}^{2/21} ({{k}_{2}^{16/21}\over k_{3}})]^{147/170}.\label{eqn.10}
\end{eqnarray}

When $t \geq {t}_{\rm gas}$, the thin disk region continues advancing inwards. In the outermost region, electron scattering opacity is gradually replaced by free-free absorption. Since the outer region is now dominated by gas pressure, the kinetic viscosity should be described by $\nu \propto {R}^{3/5}$ accordingly \citep[][]{ShenM12}, and hence its accretion rate $\propto {t}^{-19/14}$. Similarly the formulae describing the disk evolution can be derived to be,
\begin{eqnarray}
  && r_{\rm out}(t) = k_{4}(\frac{t}{t_{\rm f}})^{5/7},\nonumber\\
  && \dot{m}(r,t) =  \left\{ \begin{array}{ll}
  j_{1}(\frac{t}{t_{\rm f}})^{-19(1-s)/14}(\frac{r}{r_{\rm f}})^{s} & \mbox{if $r_{\rm i}   < r < r_{\rm tra}$},\\
  k_{5}(\frac{t}{t_{\rm f}})^{-19/14}      & \mbox{if $r_{\rm tra} < r < r_{\rm out}$},
  \end{array}
  \right.\nonumber\\
  && T(r,t) = \left\{\begin{array}{ll}
  T_{1}(\frac{\dot{m}}{\dot{m}_{0}})^{1/4}{(\frac{r}{r_{\rm f}})^{-5/8}} & \mbox{if $r_{\rm i}  < r < r_{\rm tra}$},\\
  T_{2}(\frac{r}{r_{\rm f}})^{-3/4} & \mbox{if $r_{\rm tra} < r < r_{\rm gas}$},  \\
  T_{3}(\frac{\dot{m}}{\dot{m}_{0}})^{2/5}(\frac{r}{r_{\rm f}})^{-9/10}      & \mbox{if $r_{\rm gas}   < r < r_{\rm ff}$},  \\
  T_{4}(\frac{\dot{m}}{\dot{m}_{0}})^{3/10}(\frac{r}{r_{\rm f}})^{-3/4}     & \mbox{if $r_{\rm ff}   < r < r_{\rm out}$},
  \end{array}
  \right.\nonumber\\
  && \rho(r,t) = \left\{\begin{array}{ll}
  \rho_{1}(\frac{\dot{m}}{\dot{m}_{0}})(\frac{r}{r_{\rm f}})^{-3/2}                                & \mbox{if $r_{\rm i}   < r < r_{\rm tra}$},\\
  \rho_{2}(\frac{\dot{m}}{\dot{m}_{0}})^{-2}(\frac{r}{r_{\rm f}})^{3/2}                          & \mbox{if $r_{\rm tra} < r < r_{\rm gas}$},  \\
  \rho_{3}(\frac{\dot{m}}{\dot{m}_{0}})^{2/5}(\frac{r}{r_{\rm f}})^{-33/20}      & \mbox{if $r_{\rm gas}   < r < r_{\rm ff}$},  \\
  \rho_{4}(\frac{\dot{m}}{\dot{m}_{0}})^{11/20}(\frac{r}{r_{\rm f}})^{-15/8}     & \mbox{if $r_{\rm ff}   < r < r_{\rm out}$},
  \end{array}
  \right.\label{eqn.11}
\end{eqnarray}
where the boundary $r_{\rm ff}$ separates the region dominated by electron scattering  from that  by free-free absorption \citep{ShakuraS73},
\begin{eqnarray}
  r_{\rm ff} \simeq 1.5\TDP{4}\dot{m}^{2/3},\label{eqn.12}
\end{eqnarray}
and
\begin{eqnarray}
  && j_{\rm 1}=k_{1}(\frac{t_{\rm gas}}{t_{\rm f}})^{3(1-s)/14}, \quad k_{4}=k_{3}(\frac{t_{\rm gas}}{t_{\rm f}})^{-3/7}, \quad k_{5}=k_{2}(\frac{t_{\rm gas}}{t_{\rm f}})^{3/14}, \nonumber\\
  && T_{3}=T_{2}[281 ({\alpha m\over 0.1})^{2/21}({\dot{m}_{0}^{16/21}\over r_{\rm f}})]^{3/20} (\frac{\dot{m}_{0}}{k_{2}})^{2/7} (\frac{t_{\rm gas}}{t_{\rm f}})^{16/49}, \nonumber\\
  && \rho_{3}=\rho_{2} [281 ({\alpha m\over 0.1})^{2/21}({\dot{m}_{0}^{16/21}\over r_{\rm f}})]^{63/20}, \nonumber\\
  && T_{4}=T_{3} (1.5\TDP{4}\cdot {\dot{m}_{0}^{2/3}\over r_{\rm f}})^{-3/20}, \quad \rho_{4}=\rho_{3} (1.5\TDP{4}\cdot {\dot{m}_{0}^{2/3}\over r_{\rm f}})^{9/40}.\label{eqn.13}
\end{eqnarray}
Here $\rho_{4}$ and $T_{4}$ are derived from the continuity of the
density and temperature in the outer disk at the time $t_{\rm ff}=t_{\rm
f}(1.5\TDP{4}\cdot{k}_{5}^{2/3}/k_{4})^{21/34}$ (i.e., when the
opacity of the outermost region begins to be dominated by free-free
absorption, $r_{\rm ff}=r_{\rm out}$), respectively.
When $t_{\rm gas} \leq {t} < t_{\rm ff}$, the entire disk is composed of an inner, slim disk region, an intermediate, thin disk region dominated by radiation pressure, and an outer, thin disk region dominated by gas pressure plus electron scattering opacity. When ${t} \geq {t}_{\rm ff}$, an additional, thin disk region, supported by gas pressure plus free-free absorption opacity, appears in the outermost disk region.

Phase 2 terminates when the inner, slim disk region is entirely replaced by a thin disk, i.e., $r_{\rm tra}=r_{\rm i}$,
corresponding to the time
\begin{eqnarray}
  && t_{2}=t_{\rm f}(\frac{k_{5}}{r_{\rm i}})^{14/19}.\label{eqn.14}
\end{eqnarray}

In phase 3, the whole disk is optically thick and geometrically thin,  i.e., radiative cooling becomes efficient everywhere. The disk comprises an inner region supported by radiation pressure plus electron scattering opacity, a middle region by gas pressure plus electron scattering opacity and an outer region by gas pressure plus free-free absorption opacity. The evolutionary functions in this phase are same as in Eq.~(\ref{eqn.11}) with $r>r_{\rm tra}$.

Phase 4 begins when ADAF appears in the inner disk region, or the transition radius $r_{\rm ADAF}$ between ADAF and the thin disk region equals $r_{\rm i}$. ADAF forms when radiation cooling becomes inefficient so that the heat retained in the accretion flow is mostly advected onto the central compact star. Besides the high mass accretion rate in the case of slim disks, here the very low density of the gas can also inhibit cooling when the accretion rate reaches a few percent of the Eddington accretion rate. The magnitude of $r_{\rm ADAF}$ is somewhat uncertain. It has been empirically deduced from the observations of X-ray binaries and super massive black holes in other galaxies to be \citep{YuanNara04,CaoWang14},
\begin{equation}
  r_{\rm ADAF} \simeq \dot{m}^{-1/2},\label{eqn.15}
\end{equation}
from which the end time of phase 3 is obtained to be
\begin{eqnarray}
  t_{3}=(k_{5}{r}_{\rm i}^{2})^{14/19}t_{\rm f}.\label{eqn.16}
\end{eqnarray}

Similar as in the slim disk case, we assume that there are outflows accompanied with ADAF, thus $\dot{m} \propto r^{p}$ \citep{BlandBegel99}. Based on Eq.~(\ref{eqn.15}) and the evolution of $\dot{m}$ in the thin disk region, $\dot{m}(r) \propto t^{-19/14}$, the accretion rate in the ADAF region can be written as
\begin{eqnarray}
  \dot{m}(r,t)=j_{2}(\frac{t}{t_{\rm f}})^{-19(1+p/2)/14}(\frac{r}{r_{\rm f}})^{p},\label{eqn.17}
\end{eqnarray}
where
\begin{eqnarray}
  j_{2}={r}_{\rm f}^{p}k_{5}^{1+p/2}.\label{eqn.18}
\end{eqnarray}

During the whole evolution, the temperature of the disk keeps decreasing. One can work out the time $t_{\rm crit}$, at which the disk starts to experience thermal-viscous instability, by equating the accretion rate with the critical accretion rate for different chemical compositions \citep[most likely He, C or O;][]{MenouPH02},
\begin{eqnarray}
  && \dot{m}_{\rm crit}^{\rm He}(r)=5.74\TDP{-14}(\alpha/0.1)^{0.41}{m}^{0.75}{r}^{2.62},\nonumber\\
  && \dot{m}_{\rm crit}^{\rm C}(r)=7.39\TDP{-13}(\alpha/0.1)^{0.44}{m}^{0.49}{r}^{2.23},\nonumber\\
  && \dot{m}_{\rm crit}^{\rm O}(r)=1.86\TDP{-11}(\alpha/0.1)^{0.45}{m}^{0.37}{r}^{2.05}.\label{eqn.19}
\end{eqnarray}

In the above derivations, we simply assume that the disk can expand
all the way. Actually both self-gravity and the neutral process can
pose strong constraints on the outer radius and hence the disk
evolution. If $r_{\rm sg}<r_{\rm out}$, the disk will be truncated by
self-gravity. Meanwhile, the disk becomes neutral when the local
temperature is very low. This yields the neutral radius $r_{\rm
neu}$ where $T(r_{\rm neu})=T_{\rm p}$. The outer radius of an
active disk should not be beyond either $r_{\rm sg}$ or $r_{\rm
neu}$. So we have
\begin{equation}
  r_{\rm out}= \left\{ \begin{array}{ll}
  r_{\rm out}  & {\rm if\ } r_{\rm out}<r_{\rm sg} \,\&\, r_{\rm
  neu},\\
  \min(r_{\rm sg},r_{\rm neu}) & {\rm if\ } r_{\rm out}\ge r_{\rm sg},r_{\rm
  neu}.
  \end{array}
  \right.\label{eqn.20}
\end{equation}
Note that in the presence of disk winds, the magnitude of the
accretion rate critically depends on the outer boundary. The
corresponding evolutionary functions are worked out and presented in
the Appendix \ref{seca.1} and \ref{seca.2}. In the following, we present the numerically calculated results with different input parameters.

\section{Numerical results}
\label{sec.3}

For simplicity here we assume that the Shakura \& Sunyaev viscosity
parameter is invariant (i.e., $\alpha=0.1$), and the power-law
indices related to wind loss are same, i.e., $p=s$. By changing the
initial parameters we construct several models, and allocate a name
for each model with four parameters ($\eta$, $t_{\rm f}$, $r_{\rm
i}$ and $r_{\rm f}$), {where $r_{\rm i}$ actually does not
appear unless $r_{\rm i}\neq 2.5$, neither does $r_{\rm f}$ unless $r_{\rm f}\neq 1000$}. For
example, the name $\eta0.1t_{\rm f}0.5r_{\rm i}6$ denotes the model
parameters of $\eta=$0.1, $t_{\rm f}=$0.5 s, $r_{\rm i}=6$, and
$r_{\rm f}=1000$.

According to Eqs. (\ref{eqn.1}), (\ref{eqn.3}),  (\ref{eqn.9}) and
(\ref{eqn.11}), the self-gravity radius ($r_{\rm sg}$) is a
piecewise continuous function with three turning points at
$t=t_{1}$, $t_{\rm gas}$ and $t_{\rm ff}$. Once $r_{\rm sg}$ begins
to be less than the outer radius ($r_{\rm out}$), $r_{\rm out}$ will
follow the evolution of $r_{\rm sg}$. Therefore, depending on the
evolutionary parameters at these stages ($t_{\rm f}<t<t_{1}$,
$t_{\rm 1}<t<t_{\rm gas}$, $t_{\rm gas}<t<t_{\rm ff}$, $t>t_{\rm
ff}$), the occurrence of the self-gravity truncation (i.e., the time point
when $r_{\rm out}=r_{\rm sg}$) should be estimated beforehand.
Based on the stage when the self-gravity truncation takes place for the first
time (see Appendix \ref{seca.1} for details), the evolution of the
disks with different wind loss index can be figured out.

\begin{figure}
\begin{center}
  \includegraphics[width=0.9\textwidth]{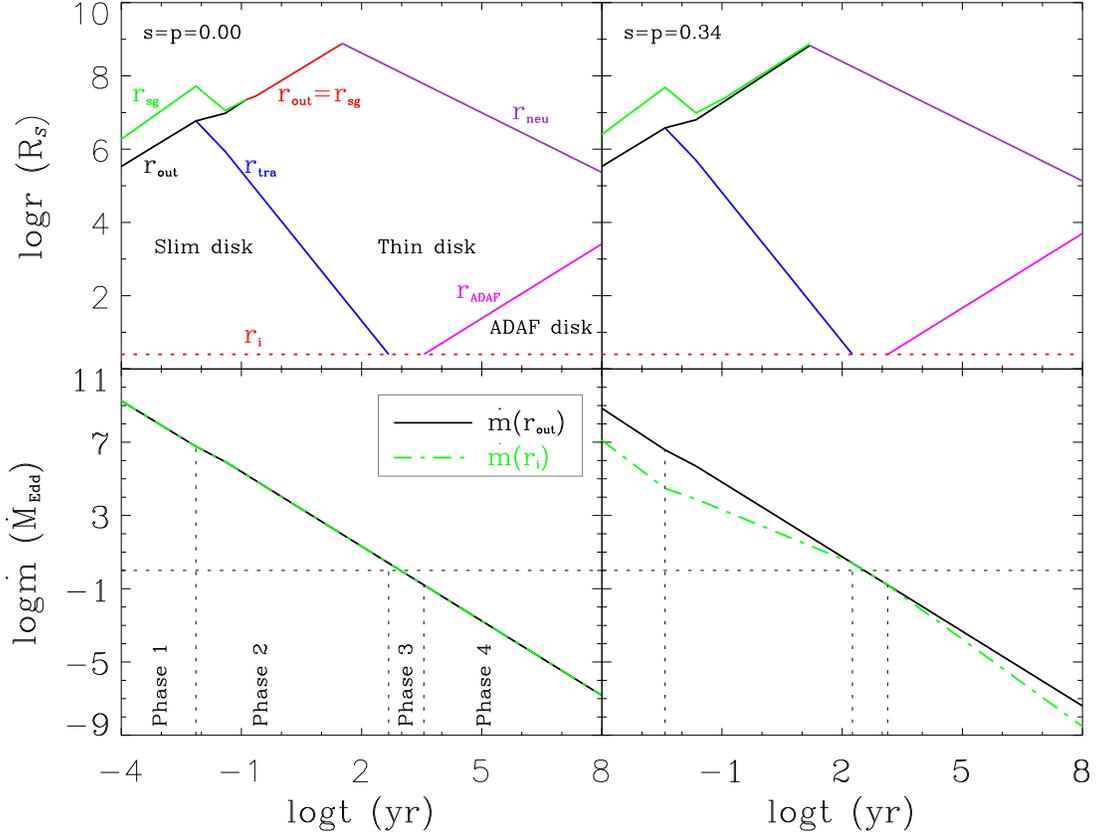}
  \caption{Evolution of the disk radii and the accretion rate at the inner region ($r_{\rm i}$)
  and the outermost region ($r_{\rm out}$) during the four phases in the model $\eta$0.1$t_{\rm f}$0.5.
  Here $r_{\rm sg}$ is the self-gravity radius, $r_{\rm neu}$ is the boundary between the cold disk with
  temperature of $\leqslant$ 300 K and the inner hot disk, $r_{\rm tra}$ is the transition radius
  between the slim disk and the thin disk, $r_{\rm ADAF}$ detaches the inner ADAF from the outer thin disk.}
  \label{fig.1}
\end{center}
\end{figure}

We regard model $\eta0.1t_{\rm f}0.5$ as our reference model. Based
on the estimation of the time $t_{\rm sg}$ when the self-gravity
truncation first occurs, we can predict the evolution of the disks
with any wind loss index ($s$). In the model, if the wind loss index
satisfies $0.19 \lesssim  {s} \lesssim 0.4$, neutralization acts in the outer disk
radius before self-gravity. In Fig.~\ref{fig.1}, the top panels
demonstrate the characteristic radii in the disks, i.e., the outer
disk radius ($r_{\rm out}$), the self-gravity radius ($r_{\rm sg}$), the
neutralization radius ($r_{\rm neu}$), the transition radii between the
slim and thin disk regions ($r_{\rm tra}$), and between ADAF and the thin
disk regions ($r_{\rm ADAF}$), as a function of time with two typical values
of $s=0$ and 0.34. The dotted horizontal line denotes the inner
radius ($r_{\rm i}$). Here the effects of the self-gravity truncation and neutral process are clearly demonstrated. While self-gravity tries to
truncate the disk, the neutral process limits the outer disk radius
within the self-gravity radius. In order to illustrate the variations of the disk structure in the four phases, we show the evolution of the
accretion rates at the inner and outer edge (denoted as
$\dot{m}(r_{\rm i})$ and $\dot{m}(r_{\rm out})$, respectively) in
the lower panels of Fig.~\ref{fig.1} (note that in \citet{ShenM12} only $\dot{m}(r_{\rm i})$ is plotted). The solid line and the green dot-dashed line denote $\dot{m}(r_{\rm out})$ and  $\dot{m}(r_{\rm in})$, respectively, which obviously evolve in different ways. As seen from the figure, the
duration of a slim disk (phase 1) is less than $0.01-0.1$ yr, during
which the accretion rate is very high (super-Eddington). In the
following $\sim 10^2-10^5$ yr (phase 2, see also Fig.~\ref{fig.3}),
the decrease in the accretion rate leads to the development of a thin
disk starting at the outer region. Subsequently the entire disk
becomes thin, and this phase (phase 3) lasts for about $10^3-10^6$
yr, until the inner region starts to become ADAF. The ADAF then
develops outwards in phase 4.

\begin{figure}
\begin{center}
  \includegraphics[width=0.9\textwidth]{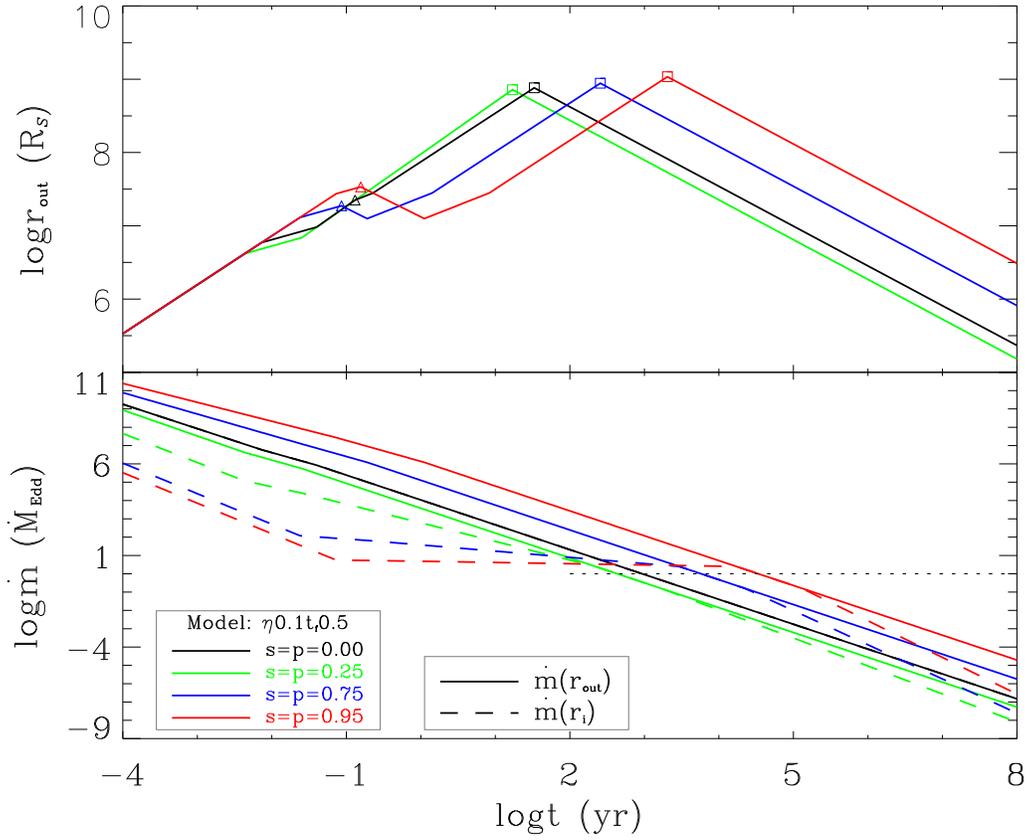}
  \caption{Evolution of the outer disk radii and the accretion rate in the model $\eta$0.1$t_{\rm f}$0.5,
  with different values of $s$. The triangle and rectangle denote the start of self-gravity truncation
  and neutralization, respectively. }
  \label{fig.2}
\end{center}
\end{figure}

\begin{figure}
\begin{center}
  \includegraphics[width=0.9\textwidth]{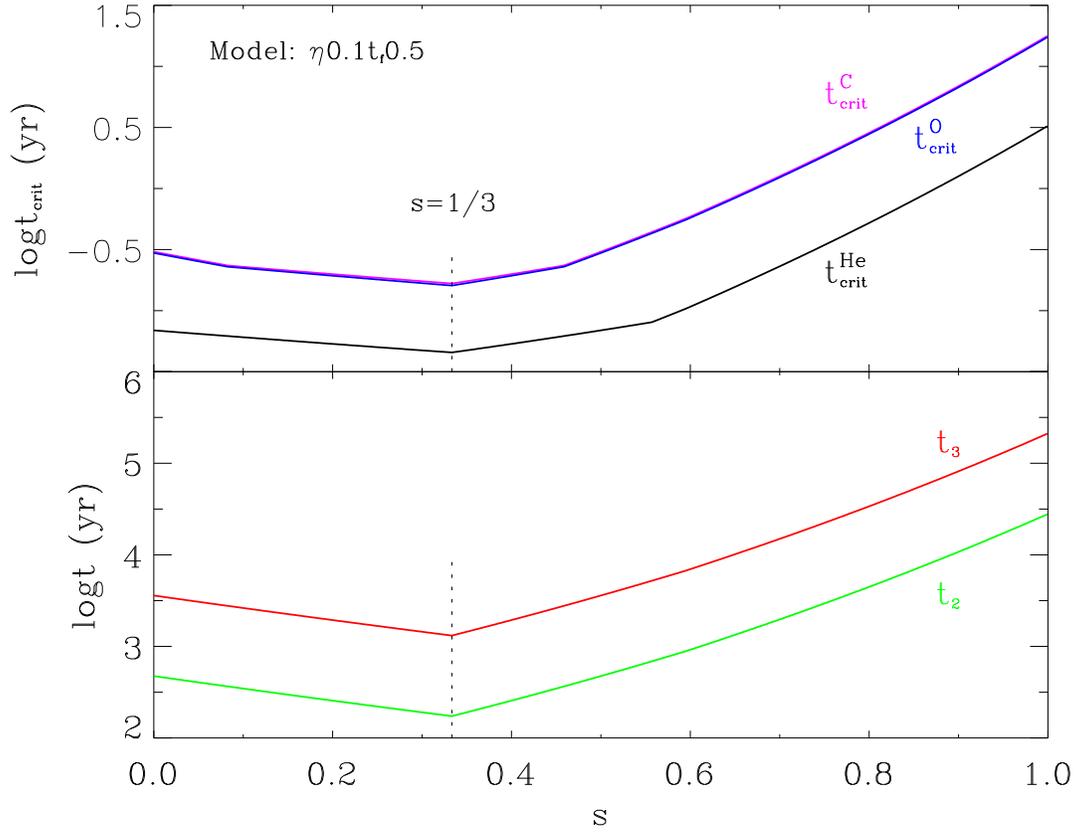}
  \caption{The upper panel shows the time when the disk becomes thermally
  unstable in the model $\eta$0.1$t_{\rm f}$0.5 as function of $s$. The bottom panel shows the times for the whole disk becoming
  optically thick ($t_{\rm 2}$) and the appearance of ADAF.}
  \label{fig.3}
\end{center}
\end{figure}

Figure~\ref{fig.2} shows the evolution of $r_{\rm out}(t)$,
$\dot{m}(r_{\rm out},\,t)$ and $\dot{m}(r_{\rm i},\,t)$  with
different $s$ values. In the case of weak (strong) disk wind, status
transitions occur earlier (later) with increasing $s$ (see the top
panel). This feature is caused by the evolution of $\dot{m}(r_{\rm
out})$, $T(r_{\rm out})$ and $\rho(r_{\rm out})$ in phase 1, which
determines the beginning time of the subsequent phase. For weak or
strong wind, $\dot{m}(r_{\rm out}) \propto t^{-(4+s)/3}$ or $\propto
t^{(2s-5)/3}$, so larger $s$ means a faster or slower decrease of
$\dot{m}$ with time, and a faster or slower status transition, respectivly (see also Fig.~\ref{fig.3}).

An important phenomenon in the evolution is the occurrence of
thermal instability in the disk. The upper panel of Fig.~\ref{fig.3}
describes how the critical timescales ($t_{\rm crit}$) evolve with
$s$ by comparing the accretion rate at $r_{\rm out}$ with the
critical accretion rate for the disk instability with different chemical
compositions (He, C and O). Since the time for the disk to be
unstable is very short ($\lesssim 20$ yr),  the accreting NS should
be observed as a very bright, persistent source only for a short time,
then quickly becomes a transient source with its peak luminosity
decreasing with time.

\begin{figure}
\begin{center}
  \includegraphics[width=0.75\textwidth]{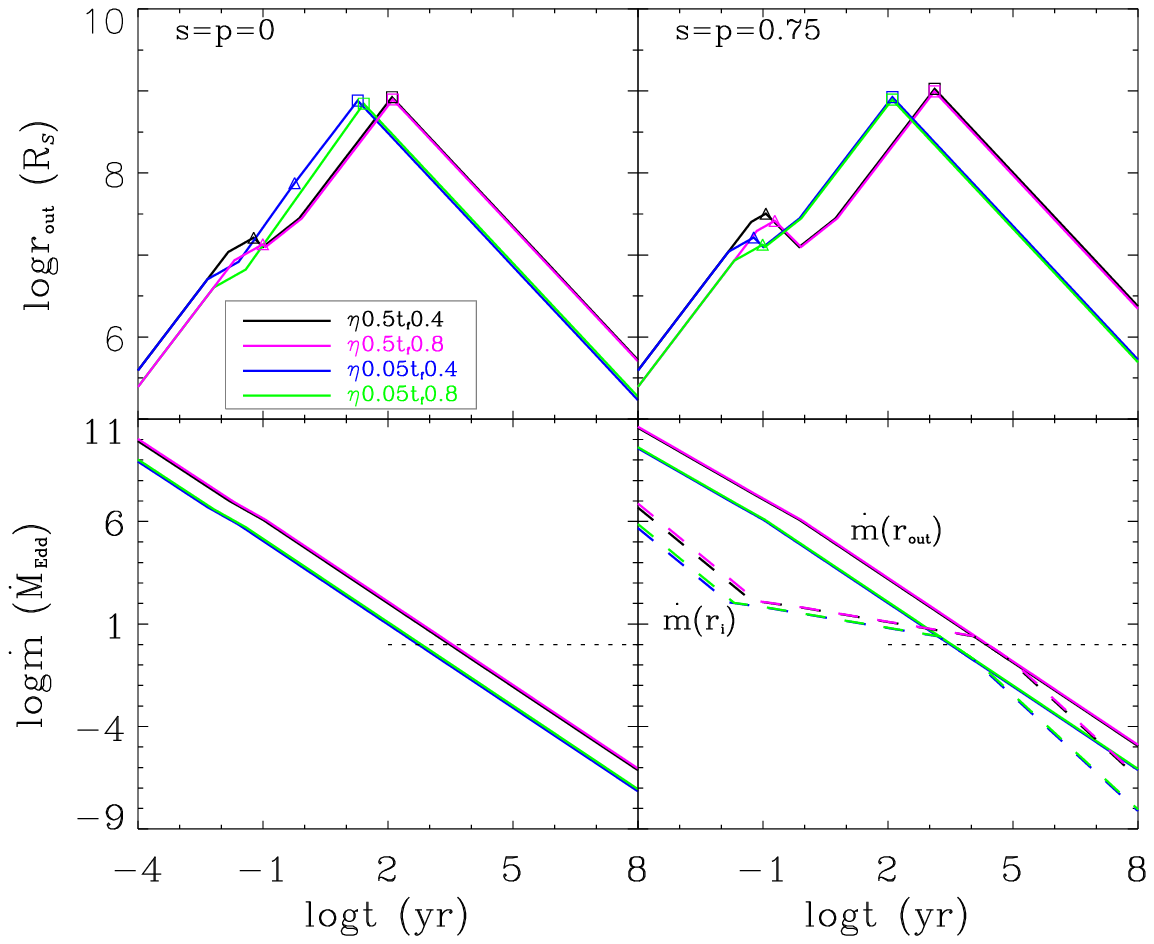}
  \includegraphics[width=0.75\textwidth]{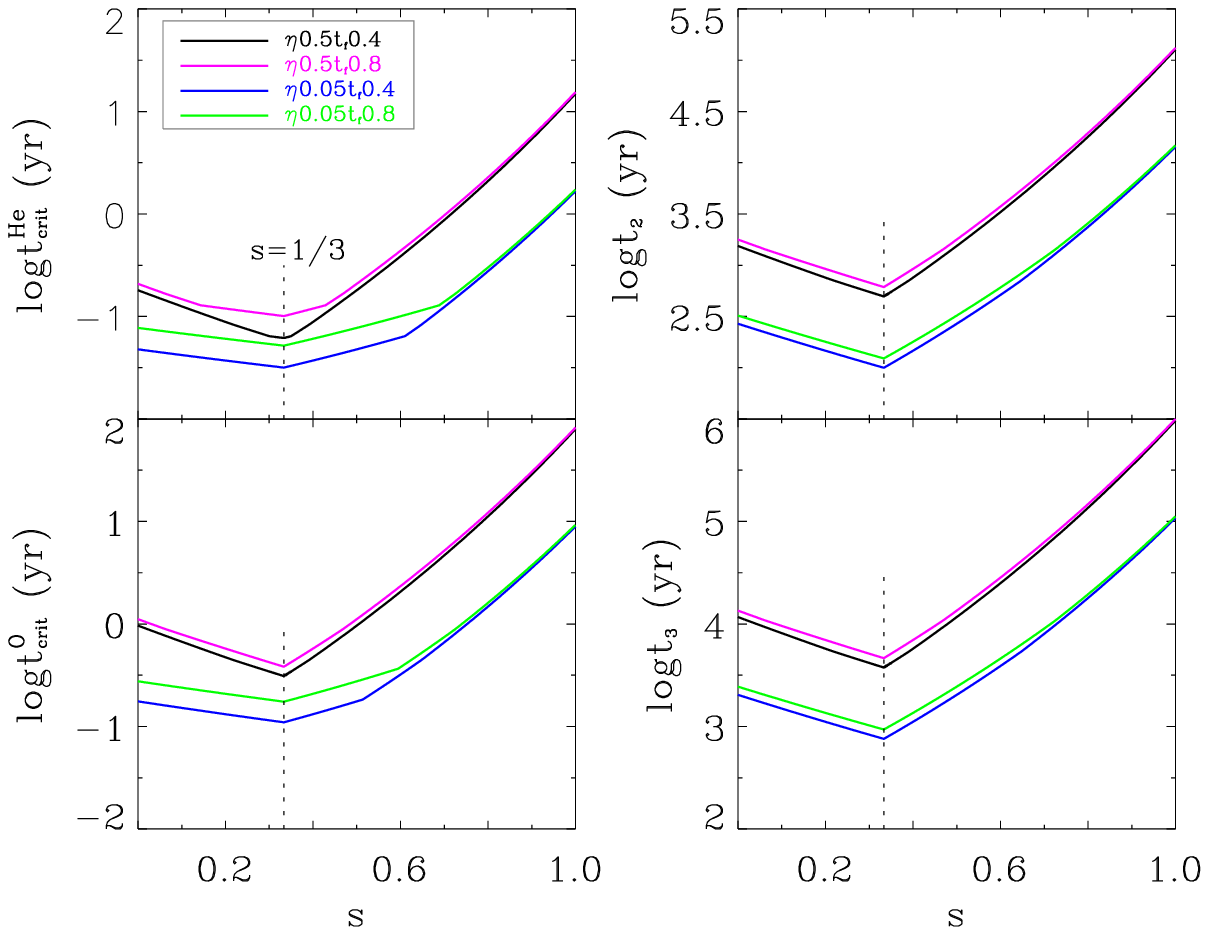}
  \caption{Models with different values of $\eta$ and $t_{\rm f}$
  are compared by simulating the evolution of accretion rate $\dot{m}(r_{\rm out})$, $\dot{m}(r_{\rm i})$,
  $r_{\rm out}$, $t_{\rm crit}$, $t_{\rm 2}$, and $t_{\rm 3}$. Here four models with $r_{\rm i}=2.5$,
  $r_{\rm f}=1000$, as labeled by legends, are considered.}
  \label{fig.4}
\end{center}
\end{figure}

Figure~\ref{fig.4} compares the evolution of the accretion rates
$\dot{m}(r_{\rm out})$ and $\dot{m}(r_{\rm i})$, the outer disk
radius $r_{\rm out}$, and the three times $t_{\rm crit}$, $t_{\rm
2}$ and $t_{\rm 3}$ with different values of  $t_{\rm f}$ and
$\eta$. Obviously the larger $\eta$, the longer the times for
status-transition (i.e., $t_{\rm crit}$, $t_{\rm 2}$ and $t_{\rm
3}$), irrespective of $s$, and the outer disk radius is more likely
to be truncated by self-gravity. Similarly, the longer $t_{\rm f}$,
usually the later the status-transitions and neutralization. In
addition, the times discussed here all have a minimum at $s \simeq
1/3$, which is caused by the $\dot{m}$ evolution in phase 1.

Finally we examine how the evolution changes with $r_{\rm i}$ and
$r_{\rm f}$. We find that, except $\dot{m}(r_{\rm i},\,t)$, neither $r_{\rm
out}(t)$ nor $\dot{m}(r_{\rm out},\,t)$ displays significant changes
with the variation in $r_{\rm i}$. The effect of the variations in $r_{\rm f}$  manifests in the evolution of $r_{\rm out}$, i.e., the
disk is more easily truncated by  self-gravity with a larger $r_{\rm f}$.

\section{Discussion}
\label{sec.4}

Although there is no direct observational evidence of the WD
merger/AIC events,  our calculations can be used to predict some
possible observational features of the post-AIC systems. The newborn
NS may firstly be observed as a super-Eddington persistent X-ray source, and quickly evolve to be transient within
$\lesssim$ 100 yr. Then the transient source becomes fainter, and
its peak X-ray luminosity during outbursts is expected to decrease
with time, until the disk becomes passive. Considering the evolution
of the mass accretion rate, the post-AIC system tends to spend a
very long time as a faint transient source. Here the transient event
originates from a similar physical process to that in low-mass X-ray
binaries (LMXBs), which occurs once the accretion rate is lower than
a critical value \citep[e.g.,][]{Lasota01}. These features may
strongly constrain the models involving fossil disk accretion onto
NSs, for example, those for anomalous X-ray pulsars and soft
$\gamma$-ray repeaters \citep[e.g.,][]{Alpar01,ErtanEE09}.

A large fraction of LMXBs are transient. They spend most  of their
lifetime at  quiescence with very low X-ray luminosities of $L_{\rm
X}\sim \DP{31-34} \es$, and are occasionally active during outbursts
with peak luminosities $L_{\rm X} \sim \DP{36-39} \es$ (in the 2-10
keV band). The typical duty cycle (DC) is $\lesssim$ 10\%
\citep[e.g.,][]{ChenSL97,WilliamsNG06}. Their transient behavior can
be explained with the thermal-viscous instability in the accretion
disk irradiated by the X-ray radiation from the central compact
object, which causes the disk to transit between a cold-neutral
state and a hot-ionized state \citep{Lasota01}.

In the last decade a new type of X-ray transients called  very faint
X-ray transients (VFXTs) were discovered. Their most striking
feature is that the $2-10$ keV luminosities during outbursts ($\sim
\DP{34-36}\es$) are at least $10-100$ lower than those of normal
X-ray transients
\citep[e.g.,][]{SakanoWD05,MunoPB05b,DegenaarWij09,DegenaarWijC12}\footnote{Some
bright accretion-disk-corona (ADC) sources appear very faint if
observed at a large orbital-inclination angle
\citep[e.g.,][]{MunoLB05a,WijnandsZR06}. Statistical analysis shows
that this kind of sources should contribute a small amount of VFXTs
rather than the bulk, so the sub-luminous outbursts are intrinsic
\citep{KingW06,DegenaarWijC12}.}. However, this classification
sensitively depends on the distance determinations of the sources
\citep{ArmasWD14}. In addition there are hybrid systems displaying
both normal and very faint outbursts
\citep{WijnandsMW02,MunoBA03,DegenaarWij09,DelSantoSR10,DegenaarWij10}.

A large fraction of known VFXTs are very close to Sgr A*
\citep[within $10'$,
e.g.,][]{MunoPB05b,PGB05,WijnandsZR06,DegenaarWij09} and only a
handful sources are observed far from it
\citep{HandWW04,HeinkeCL09,HeinkeAlta10,BozzoFSB11}, which might be
due to the preferred high-resolution X-ray observations in the
Galactic center \citep{DegenaarWij09}. Most known VFXTs exhibit high
Galactic absorption, with only two exceptions \citep[M15 X-3 and
Swift J1357.2-0933,][]{HeinkeCL09,ArmasDRW13}. The optical
companions of most VFXTs must be fainter than B2 V stars, since no
infrared or optical counterparts with a magnitude of $K<15$ have so
far been detected \citep{MunoPB05b,MauerhanM09,DegenaarWijC12}.

The vast majority of VFXTs are likely to be accreting NS  and BH
systems since only one accreting WD has exhibited outbursts with
peak luminosities above $\DP{34} \es$ \citep[i.e., GK
Per,][]{WatsonKO85}. Observations of VFXTs suggest that they should
comprise some heterogeneous populations
\citep[e.g.,][]{MunoPB05b,SakanoWD05,KingW06,DegenaarWij09,ArmasDP11,DegenaarWijC12}.
By analogy with the normal transients, if the typical
radiation-efficiency of accretion ($\sim \DP{20}$ erg/g) and DC
($\lesssim$ 10\%) are applied to VFXTs, the long-term time-averaged
accretion rate $\langle\dot{M}\rangle$ should be $\lesssim
\DP{-13}-\DP{-12} \SMPY$ \citep{KingW06,HeinkeBDW15}. LMXBs with
such low $\dot{M}$ can hardly evolve from the traditional Roche-lobe
overflow (RLOF) process within Hubble time, unless some abrupt
change of $\dot{M}$ occurs \citep[e.g.,][]{MaccaroneP13}.

\citet{KingW06} proposed that part of the VFXTs may be  LMXBs
harboring either an extremely low-mass companion (such as a brown
dwarf or a planet) or an intermediate-mass BH to guarantee that the
average mass transfer rate is extremely low, but the birth rates of
such systems are highly uncertain.

The propeller effect of a NS accretor may also lead to a
particularly low accretion rate
\citep{DegenaarWij09,HeinkeCL09,ArmasDP11}. Accretion onto a NS
can be inhibited by the centrifugal barrier unless its spin angular
velocity is smaller than the Keplerian angular velocity at the
magnetosphere \citep{IllarionovS75}. Numerical simulations show that
diffusive and viscous interaction between the magnetosphere and the
disk may play an important role in the propeller-driven outflows and
there is always a slow accretion flow in the case of weak-propeller
systems \citep{RomanovaUK04,RomanovaUK05}. Based on this picture,
\citet{DegenaarWijR14} tried to explain the very faint persistent
outburst of XMM J174457$-$2850.3. This model can be tested by
future measurements of the spin and the magnetic field of the
accreting NS.

Here we suggest that an accreting NS (formed via WD merger) from a
fossil disk may account for part of the VFXT population, considering
the fact that $\gtrsim$ 1/3 VFXTs are NS systems and the optical
counterparts, if exist, are either very faint or undetectable.
Whether the merger remnant can evolve into a VFXT depends on the
evolution of the fossil disk, provided that the NS has a weak
magnetic field ($<10^{10}$ G). The key points are whether the disk
can be thermally unstable and maintain a very low accretion rate
$\lesssim \DP{-12}\SMPY$ for a sufficiently long time. The time when
the system reaches such low accretion rate is an increasing function
of the initial disk mass (or its formation timescale), and is
estimated to be $\gtrsim \DP{5}$ yr from our previous calculations.

Population synthesis calculations \citep[e.g.,][]{CJZH12}  show that
the Galactic birthrate of NSs born via double WD mergers is about
$5\times10^{-4}$ yr$^{-1}$. Adopting a lifetime of $\gtrsim 10^5$ yr,
we can then roughly estimate that there might be $\gtrsim$ 100 transient
NSs accreting from a fossil disk, with the properties resembling
VFXTs. For such objects, some observational features can distinguish
them from those in LMXBs:

(i)  No orbital modulations would be displayed in its light curve
since there is not any companion star.

(ii) One may expect to detect spectral lines from C or O
elements in the disk.

(iii) The delay time for WD mergers can be as short as a few $10^8$
yr \citep{CJZH12}, so WD mergers may trace the star-formation
processes. This may partly explain why there are a rich population
of VFXTs in the Galactic center region.

\section{Conclusions}
\label{sec.5}

AIC is expected to be an important channel to the formation of MSPs
theoretically \citep[e.g.,][]{HurleyTW10}, although direct
observational evidence is still lacking. In this paper we
investigate the evolution of the merger remnant of double WDs (i.e.,
a fossil disk around a NS), taking into account the effect of the
outer radius, which is determined by either self-gravity or
neutralization.

We find that the evolution generally consists of four phases. (i) In
phase 1, the newborn fossil disk is a slim one, and accretes onto
the NS at a super-Eddington rate. As a consequence, the system
should be observed as a very bright persistent source. However, the
duration of the phase is less than 1 yr. (ii) In phase 2, the slim
disk gradually evolves to be a thin disk with the decrease of the
accretion rate, since radiative cooling becomes effective. The
transition starts from the outer region and develops inwards until
the entire disk turns into a thin disk (within a time $\sim
\DP{1-6}$ yr). During the evolution, the disk is expected to be
thermally unstable within $\lesssim$ 100 yr, i.e., the system changes to
a bright transient X-ray source, and the peak X-ray luminosity
decreases with time. The outer disk either is truncated by self-gravity or becomes neutral and stops spreading any more. (iii) In
phase 3 the whole disk is a thin disk. The evolution of the thin
disk can last $\sim \DP{2-7}$ yr, during which the X-ray
brightness of the transient source becomes even lower. (iv) In phase
4, the thin disk turns into an ADAF from the inner region once the
accretion rate decrease to $\lesssim 0.01 \dot{M}_{\rm Edd}$, where
advective cooling is dominated.

During the disk evolution, we consider possible wind loss since the
accretion rate can be significantly higher or lower than
$\dot{M}_{\rm Edd}$ \citep{BlandBegel99,NarayanPK01}. In our
calculations, the outer radius of the disk is firstly determined.
Based on the estimation, the evolution of the disk with different wind
loss indices can then be predicted. One can estimate when the disk
evolves from a slim disk to a thin disk, and when the NS evolves
from persistent to transient, depending on the self-gravity truncation
or the neutral process. On
one hand, the disk will spread more slowly if truncated by the self-gravity. On the other hand, the self-gravity truncation tends to postpone the
occurrence of the neutralization.

Based on the calculations, a possible picture of the fossil disk
accretion onto an isolated NS is predicted. Unlike previous studies
on the early evolution of the fallback disk, here we are interested
in the evolution of the post-AIC systems at the late phase, during
which the source appears faint ($L_{\rm X}<$ {\DP{36}\es}). Indeed,
$\sim$ 40 peculiar transients, called VFXTs, have been captured by
the high-resolution monitors during the past decade
\citep[e.g.,][]{KingW06,DegenaarWij09,DegenaarWijC12}. Our
calculations demonstrate that the post-AIC systems may be transient
within 100 yr and reach a very low accretion rate of {\DP{-12}\SMPY}
at a time $\gtrsim$ {\DP{5}} yr, implying that the post-AIC systems
could be the progenitors  of some VFXTs.

\begin{acknowledgements}

We are grateful to an anonymous for clarifying and helpful comments. This work was supported by the Natural Science Foundation of China
under grant numbers 11133001 and 11333004, and the Strategic
Priority Research Program of CAS (under grant number XDB09000000).

\end{acknowledgements}

\clearpage

\appendix

If both the self-gravity truncation and the neutral process are considered, the disk evolution will behave differently, and will be described in the appendix. For simplification, the density ($\rho_{\rm i}$, i=0, 1, 2, 3, 4) in the following discussions is assumed to be nondimensional.

\section{Disk Evolution with self-gravity truncation}
\label{seca.1}

Since the disk in each phase follows the distinct evolution (Section
\ref{sec.2.2}), the self-gravity truncation can hardly be described by a
common function. If there is self-gravity truncation, we assume that
$r_{\rm out}=r_{\rm sg}$ is satisfied unless there is a new
state-transition. Depending on the initial parameters, self-gravity
truncation can occur at any stage. In addition, if the wind loss
index satisfies $s \rightarrow$1, the outer region will be easily
truncated.

Moreover, the disk may be truncated at the beginning of phase 1
under specific situations. Firstly, a timescale $t_{\rm sg0}$ when
$r_{\rm out}=r_{\rm sg}$, using the evolutionary parameters in
phase 1, is calculated to be,
\begin{eqnarray}
  && t_{\rm sg0}=\left\{ \begin{array}{ll}
  t_{\rm f}\cdot(\frac{2.92\TDP{5}}{r_{\rm f}\cdot\rho_{0}^{1/3}\cdot{m}^{2/3}})^{-9/(1+s)}  & (0<s<1/3),\\
  t_{\rm f}\cdot(\frac{2.92\TDP{5}}{r_{\rm f}\cdot\rho_{0}^{1/3}\cdot{m}^{2/3}})^{-9/(2-2s)}  & (1/3<s<1),
  \end{array}
  \right.\label{eqna01}
\end{eqnarray}
From Eq.~(\ref{eqn.3}), one can infer that increase of $r_{\rm sg}$
with time ($\sim \propto {t}^{7/9}$, $0<s<1$) is faster than that of
$r_{\rm out}$ with time ($\propto {t}^{2/3}$). Then the disk
evolution should be described by $r_{\rm out}=r_{\rm sg}$ from the
time point of $t_{\rm f}$ to that of $t=\min(t_{\rm sg0}, t_{\rm
1})$. In the following, we only consider models with $t_{\rm sg0} \leq
t_{\rm f}$. Depending on the occurrence of self-gravity truncation, the
disk evolution can be classified as follows.

(i) If the disk is truncated at $t_{1}<t<t_{\rm gas}$, the outer
radius evolves in the following laws,
\begin{eqnarray}
  && r_{\rm out}(t_{1}<t<t_{\rm sg})=k_{3}\cdot(\frac{t}{t_{\rm f}})^{2/7},\label{eqna02}\\
  && r_{\rm out}(t_{\rm sg}<t<t_{\rm gas})=r_{\rm f}\cdot(\frac{ 2.92\TDP{5}}{r_{\rm f}\cdot\rho_{2}^{1/3}\cdot{m}^{2/3}})^{2/3}\cdot (\frac{k_{2}}{\dot{m}_{0}})^{4/9}\cdot (\frac{t}{t_{\rm f}})^{-32/63},\label{eqna03}
\end{eqnarray}
where
\begin{eqnarray}
  && t_{\rm sg}=t_{\rm f}\cdot(\frac{r_{\rm f}}{k_{\rm 3}})^{63/50}\cdot(\frac{2.92\TDP{5}}{r_{\rm f}\cdot\rho_{2}^{1/3}\cdot{m}^{2/3}})^{21/25}\cdot (\frac{k_{2}}{\dot{m}_{0}})^{14/25},\nonumber\\
  && t_{\rm gas}=t_{\rm f}\cdot[\frac{281}{{r}_{\rm f}}\cdot(m\alpha/0.1)^{2/21}\cdot{k}_{2}^{16/21}\cdot (\frac{\dot{m}_{0}}{k_{2}})^{4/9}\cdot (\frac{r_{\rm f}\cdot\rho_{2}^{1/3}\cdot{m}^{2/3}}{ 2.92\TDP{5}})^{2/3}]^{441/160}.\label{eqna04}
\end{eqnarray}

When $t_{\rm gas}<t<t_{\rm 2}$, we have
\begin{eqnarray}
  && r_{\rm out}(t_{\rm gas}<t<t_{\rm ff})=r_{\rm f}\cdot(\frac{ 2.92\TDP{5}}{r_{\rm f}\cdot\rho_{3}^{1/3}\cdot{m}^{2/3}})^{20/9}\cdot (\frac{k_{5}}{\dot{m}_{0}})^{-8/27}\cdot (\frac{t}{t_{\rm f}})^{76/189},\nonumber\\
  && r_{\rm out}(t>t_{\rm ff})=r_{\rm f}\cdot(\frac{2.92\TDP{5}}{r_{\rm f}\cdot\rho_{4}^{1/3}\cdot{m}^{2/3}})^{8/3}\cdot (\frac{k_{5}}{\dot{m}_{0}})^{-22/45}\cdot (\frac{t}{t_{\rm f}})^{209/315}, \label{eqna05}
\end{eqnarray}
where
\begin{eqnarray}
  && t_{\rm ff}=t_{\rm f}\cdot[\frac{1.5\TDP{4}}{{r}_{\rm f}}\cdot {\dot{m}_{0}^{2/3}}\cdot (\frac{k_{5}}{\dot{m}_{0}})^{26/27}\cdot (\frac{r_{\rm f}\cdot\rho_{3}^{1/3}\cdot{m}^{2/3}}{ 2.92\TDP{5}})^{20/9}]^{189/247},\label{eqna06}
\end{eqnarray}
and some other parameters are referred to Eq.~(\ref{eqn.11})$-$(\ref{eqn.18}).

(ii) If self-gravity truncation occurs at $t_{\rm gas}<t<t_{\rm ff}$, the
evolution of the outer radius follows,
\begin{eqnarray}
  && r_{\rm out}(t_{\rm gas}<t<t_{\rm sg})=k_{4}\cdot(\frac{t}{t_{\rm f}})^{5/7},\nonumber\\
  && r_{\rm out}(t_{\rm sg}<t<t_{\rm ff})=r_{\rm f}\cdot(\frac{ 2.92\TDP{5}}{r_{\rm f}\cdot\rho_{3}^{1/3}\cdot{m}^{2/3}})^{20/9}\cdot (\frac{k_{5}}{\dot{m}_{0}})^{-8/27}\cdot (\frac{t}{t_{\rm f}})^{76/189},\label{eqna07}
\end{eqnarray}
where
\begin{eqnarray}
  && t_{\rm sg}=t_{\rm f} \cdot [\frac{r_{\rm f}}{k_{4}} \cdot(\frac{ 2.92\TDP{5}}{r_{\rm f}\cdot\rho_{3}^{1/3}\cdot{m}^{2/3}})^{20/9}\cdot (\frac{k_{5}}{\dot{m}_{0}})^{-8/27}]^{189/59}.\label{eqna08}
\end{eqnarray}
The evolution after $t_{\rm ff}$ can be described by
Eq.~(\ref{eqna05}) and (\ref{eqna06}).

(iii) In other cases the evolution of the outer radius is given by
\begin{eqnarray}
  && r_{\rm out}(t_{\rm ff}<t<t_{\rm sg})=k_{4}\cdot(\frac{t}{t_{\rm f}})^{5/7},\nonumber\\
  && r_{\rm out}(t>t_{\rm sg})=r_{\rm f}\cdot(\frac{ 2.92\TDP{5}}{r_{\rm f}\cdot\rho_{4}^{1/3}\cdot{m}^{2/3}})^{8/3}\cdot (\frac{k_{5}}{\dot{m}_{0}})^{-22/45}\cdot (\frac{t}{t_{\rm f}})^{209/315},\label{eqna09}
\end{eqnarray}
where
\begin{eqnarray}
  && t_{\rm sg}=t_{\rm f} \cdot [\frac{r_{\rm f}}{k_{4}} \cdot(\frac{ 2.92\TDP{5}}{r_{\rm f}\cdot\rho_{4}^{1/3}\cdot{m}^{2/3}})^{8/3}\cdot (\frac{k_{5}}{\dot{m}_{0}})^{-22/45}]^{315/16},\label{eqna10}
\end{eqnarray}
and the following evolution is similar to that described by
case~(i).

\section{Disk Evolution with Neutralization}
\label{seca.2}

In the above calculations, the effect of the neutral process is
ignored. According to our calculations, the time ($t_{\rm neu}$) for
the outer disk to become neutral satisfies $t_{\rm 2}> t_{\rm neu}>
t_{\rm ff}$. Thus, when $t > t_{\rm neu}$, the disk is composed of a
hot region ($r_{\rm i} < r < r_{\rm neu}$) and a cold region $r_{\rm
neu} < r < r_{\rm out}$, where
\begin{eqnarray}
  && r_{\rm neu}(t)=r_{\rm f}\cdot (\frac{T_{4}}{\rm 300\,\,K})^{4/3}\cdot (\frac{k_{5}}{\dot{m}_{0}})^{2/5}\cdot (\frac{t}{t_{\rm f}})^{-19/35}.\label{eqna11}
\end{eqnarray}

And the neutralization time $t_{\rm neu}$ can be calculated from the
condition $r_{\rm neu}=r_{\rm out}$, i.e.,
\begin{eqnarray}
  && t_{\rm neu}=\left\{ \begin{array}{ll}
  t_{\rm f}\cdot (\frac{T_{4}}{\rm 300\,\,K})^{21/19}\cdot (\frac{k_{5}\cdot\rho_{4}}{\dot{m}_{0}})^{14/19}\cdot (\frac{ 2.92\TDP{5}}{r_{\rm f}\cdot{m}^{2/3}})^{-42/19}  & (t_{\rm neu}>t_{\rm sg}),\\
  t_{\rm f}\cdot(\frac{T_{4}}{\rm 300\,\,K})^{35/33}\cdot (\frac{k_{5}}{\dot{m}_{0}})^{7/22}\cdot (\frac{k_{4}}{r_{\rm f}})^{-35/44}  & ({\rm otherwise}),
  \end{array}
  \right.\label{eqna12}
\end{eqnarray}
where $r_{\rm out}$ is described by Eq.~(\ref{eqna09}),  based on
whether the outer region has been truncated or not before being
neutral. Substituting Eq.~(\ref{eqna12}) into Eq.~(\ref{eqna11}),
one can obtain the maximal $r_{\rm out}$, $r_{\rm out,max}=r_{\rm
neu}(t_{\rm neu})$.

\clearpage
\bibliographystyle{apj}

\begin{thebibliography}{28}

\bibitem[{Abramowicz} et al.(1988)]{AbramowiczCLS88} Abramowicz, M. A., Czerny, B., Lasota, J. P., \& Szuszkiewicz, E. 1988, ApJ, 332, 646
\bibitem[{Aggarwal \& Oberbeck}(1974)]{AggOb74} Aggarwal, H. R., \& Oberbeck, V. R. 1974, ApJ, 191, 577
\bibitem[{Alpar}(2001)]{Alpar01} Alpar, M. A. 2001, ApJ, 554, 1245
\bibitem[{Alpar} et al.(2013)]{AlparCE13} Alpar, M. A., \c{C}al\i\c{s}kan, \c{S}., \& Ertan, \"{U}. 2013, IAUS, 290, 93
\bibitem[{Armas Padilla} et al.(2011)]{ArmasDP11} Armas Padilla, M., Degenaar, N., Patruno, A., Russell, D. M., Linares, M., Maccarone, T. J., Homan, J., \& Wijnands, R. 2011, MNRAS, 417, 659
\bibitem[{Armas Padilla} et al.(2013)]{ArmasDRW13} Armas Padilla, M., Degenaar, N., Russell, D. M., \& Wijnands, R. 2013, MNRAS, 428, 3083
\bibitem[{Armas Padilla} et al.(2014)]{ArmasWD14} Armas Padilla, M., Wijnands, R., Degenaar, N., et al. 2014, MNRAS, 444, 902
\bibitem[{Benz} et al.(1990)]{BenzBCP90} Benz, W., Bowers, R. L., Cameron, A. G. W., \& Press, W. H. 1990, ApJ, 348, 647
\bibitem[{Bhattacharya \& van den Heuvel}(1991)]{BhattaHeu91} Bhattacharya, D., \& van den Heuvel, E. P. J. 1991, Phys. Rep., 203, 1
\bibitem[{Blandford \& Begelman}(1999)]{BlandBegel99} Blandford, R. D., \& Begelman, M. C. 1999, MNRAS, 303, L1
\bibitem[{Bozzo} et al.(2011)]{BozzoFSB11} Bozzo, E., Ferrigno, C., Stevens, J., Belloni, T. M., Rodriguez, J., den Hartog, P. R., Papitto, A., et al. 2011, AA, 535, L1
\bibitem[{Burderi} et al.(1998)]{BurderiKS98} Burderi, L., King, A. R., \& Szuszkiewicz, E. 1998, ApJ, 509, 85
\bibitem[{Cannizzo \& Gehrels}(2009)]{CannizzoGeh09} Cannizzo, J. K., \& Gehrels, N. 2009, ApJ, 700, 1047
\bibitem[{Cannizzo} et al.(1990)]{CannizzoLee90} Cannizzo, J. K., Lee, H. M., \& Goodman, J. 1990, ApJ, 351, 38
\bibitem[{Cao \& Wang}(2014)]{CaoWang14} Cao, X., \& Wang, J.-X. 2014, arXiv: 1406.6442
\bibitem[{Chen} et al.(2012)]{CJZH12} Chen, X., Jeffery, C. S., Zhang, X., \& Han, Z., 2012, ApJL, 755, L9
\bibitem[{Chen} et al.(1997)]{ChenSL97} Chen, W., Shrader, C. R., \& Livio, M. 1997, ApJ, 491, 312
\bibitem[{Chevalier}(1989)]{Chevalier89} Chevalier, R. A. 1989, ApJ, 346, 847
\bibitem[{Dan} et al.(2011)]{DanRoss11} Dan, M., Rosswog, S., Guillochon, J., \& Ramirez-Ruiz, E. 2011, ApJ, 737, 89
\bibitem[{Darbha} et al.(2010)]{DarbhaMQ10} Darbha, S., Metzger, B. D., Quataert, E., Kasen, D., Nugent, P., \& Thomas, R. 2010, MNRAS, 409, 846
\bibitem[{Degenaar \& Wijnands}(2009)]{DegenaarWij09} Degenaar, N., \& Wijnands, R. 2009, AA, 495, 547
\bibitem[{Degenaar \& Wijnands}(2010)]{DegenaarWij10} Degenaar, N., \& Wijnands, R. 2010, AA, 524, A69
\bibitem[{Degenaar} et al.(2012)]{DegenaarWijC12} Degenaar, N., Wijnands, R., Cackett, E. M., Homan, J., in't Zand, J. J. M., Kuulkers, E., et al. 2012, AA, 545, 49
\bibitem[{Degenaar} et al.(2014)]{DegenaarWijR14} Degenaar, N., Wijnands, R., Reynolds, M. T., Miller, J. M., \& Altamirano, D. 2014, ApJ, 792, 109
\bibitem[{Del Santo} et al.(2010)]{DelSantoSR10} Del Santo, M., Sidoli, L., Romano, P., Bazzano, A., Wijnands, R., Degenaar, N., \& Mereghetti, S. 2010, MNRAS, 403, L89
\bibitem[{Dessart} et al.(2007)]{DessartBLO07} Dessart, L., Burrows, A., Livne, E., \& Ott, C. D. 2007, ApJ, 669, 585
\bibitem[{Dessart} et al.(2006)]{DessartBOL06} Dessart, L., Burrows, A., Ott, C. D., Livne, E., Yoon, S.-Y., \& Langer, N. 2006, ApJ, 644, 1063
\bibitem[{D'Souza} et al.(2006)]{SouzaMTF06} D'Souza, M. C. R., Motl, P. M., Tohline, J. E., \& Frank, J. 2006, ApJ, 643, 381
\bibitem[{Duncan \& Thompson}(1992)]{DuncanT92} Duncan, R. C., \& Thompson, C. 1992, ApJ, 392, L9
\bibitem[{Ertan} et al.(2009)]{ErtanEE09} Ertan \"{U}., Ek\c{s}i, K. Y., Erkut, M. H., \& Alpar, M. A. 2009, ApJ, 702, 1309
\bibitem[{Fryer} et al.(2002)]{FryerHH02} Fryer, C. L., Holz, D. E., \& Hughes, S. A. 2002, ApJ, 565, 430
\bibitem[{Garc{\'\i}a-Berro} et al.(2012)]{Garcia12} Garc{\'\i}a-Berro, E., et al. 2012, ApJ, 749, 25
\bibitem[{Garc{\'\i}a-Berro} et al.(2007)]{GarciaL07} Garc{\'\i}a-Berro, E., Lor$\acute{\rm e}$n-Aguilar, P., Pedemonte, A. G., Isern, J., et al. 2007, ApJ, 661, L179
\bibitem[{Hands} et al.(2004)]{HandWW04} Hands, A. D. P., Warwick, R. S., Watson, M. G., \& Helfand, D. J. 2004, MNRAS, 351, 31
\bibitem[{Han} et al.(2002)]{HanPod02} Han, Z., Podsiadlowski Ph., Maxted, P. F. L., Marsh, T. R., \& Ivanova, N. 2002, MNRAS, 336, 449
\bibitem[{Heber}(2009)]{Heber09} Heber, U. 2009, ARAA, 47, 211
\bibitem[{Heinke} et al.(2010)]{HeinkeAlta10} Heinke, C. O., Altamirano, D., Cohn, H. N., Lugger, P. M., Budac, S. A., Servillat, M., Linares, M., et al. 2010, ApJ, 714, 894
\bibitem[{Heinke} et al.(2015)]{HeinkeBDW15} Heinke, C. O., Bahramian, A., Degenaar, N., \& Wijnands, R. 2015, MNRAS, 447, 3034
\bibitem[{Heinke} et al.(2009)]{HeinkeCL09} Heinke, C. O., Cohn, H. N., \& Lugger, P. M. 2009, ApJ, 692, 584
\bibitem[{Holberg} et al.(2008)]{HolbergSO08} Holberg, J. B., Sion, E. M., Oswalt, T., McCook, G. P., Foran, S., \& Subasavage, J. P. 2008, AJ, 135, 1225
\bibitem[{Hurley} et al.(2010)]{HurleyTW10} Hurley, J. R., Tout, C. A., Wickramasinghe, D. T., Ferrario, L., \& Kiel, P. D. 2010, MNRAS, 402, 1437
\bibitem[{Iben \& Tutukov}(1984)]{IbenT84} Iben, I., \& Tutukov, A. V. 1984, ApJS, 54, 335
\bibitem[{Illarionov \& Sunyaev}(1975)]{IllarionovS75} Illarionov, A. F., \& Sunyaev, R. A. 1975, AA, 39, 185
\bibitem[{Inotsuka \& Sano}(2005)]{InotsukaS05} Inotsuka, S., \& Sano, T. 2005, ApJ, 628, L155
\bibitem[{Jeffery} et al.(2011)]{JefferyKS11} Jeffery, C. S., Karakas, A. I., \& Saio, H. 2011, MNRAS, 414, 3599
\bibitem[{Kato} et al.(1998)]{Kato98}  Kato, S., Fukue, J., \& Mineshige, S. 1998, Black hole accretion disks (Kyoto Univ. Press)
\bibitem[{Kawai} et al.(1987)]{KawaiSN87} Kawai, Y., Saio, H., \& Nomoto, K. 1987, ApJ, 315, 229
\bibitem[{K\"{u}lebi} et al.(2013)]{KulebiE13} K\"ulebi, B., Ek{\c s}i, K. Y., Lor\'en-Aguilar, P., Isern, J., \& Garc{\'\i}a-Berro, E. 2013, MNRAS, 431, 2778
\bibitem[{Kilic} et al.(2014)]{KilicBG14} Kilic, M., Brown, W. R., Gianninas, A., Hermes, J. J., Allende Prieto, C., \& Kenyon, S. J. 2014, MNRASL, 444, L1
\bibitem[{King} et al.(2001)]{KingPW01} King, A. R., Pringle, J. E., \& Wickramasinghe, D. T. 2001, MNRAS, 320, L45
\bibitem[{King \& Wijnands}(2006)]{KingW06} King, A. R., \& Wijnands, R. 2006, MNRAS, 366, L31
\bibitem[{Lasota}(2001)]{Lasota01} Lasota, J.-P. 2001, New Astron. Rev., 45, 449
\bibitem[{Levan} et al.(2006)]{LevanWC06} Levan, A. J., Wynn, G. A., Chapman, R., et al. 2006, MNRAS, 368, L1
\bibitem[{Liebert} et al.(2005)]{LiebertBH05} Liebert, J., Bergeron, P., \& Holberg, J. B. 2005, ApJS, 156, 47
\bibitem[{Longland} et al.(2011)]{LonglandLJ11} Longland, R., Lor\'en-Aguilar, P., Jos$\acute{\rm e}$ J., Garc{\'\i}a-Berro, E., et al. 2011, ApJL, 737, L34
\bibitem[{Lor\'en-Aguilar} et al.(2009)]{LorenIsern09} Lor\'en-Aguilar, P., Isern, J., \& Garc{\'\i}a-Berro, E. 2009, AA, 500, 1193
\bibitem[{Lynden-Bell \& Pringle}(1974)]{LyndenP74} Lynden-Bell, D., \& Pringle, J. E. 1974, MNRAS, 168, 603
\bibitem[{Maccarone \& Patruno}(2013)]{MaccaroneP13} Maccarone, T. J., \& Patruno, A. 2013, MNRAS, 428, 1335
\bibitem[{Mauerhan} et al.(2009)]{MauerhanM09} Mauerhan, J. C., Muno, M. P., Morris, M. R., et al. 2009, ApJ, 703, 30
\bibitem[{Menou} et al.(2001)]{MenouPH01} Menou, K., Perna, R., \& Hernquist, L. 2001, ApJ, 559, 1032
\bibitem[{Menou} et al.(2002)]{MenouPH02} Menou, K., Perna, R., \& Hernquist, L. 2002, ApJ, 564, L81
\bibitem[{Metzger} et al.(2008)]{MetzgerPQ08} Metzger, B. D., Piro, A. L., \& Quataert, E. 2008, MNRAS, 390, 781
\bibitem[{Metzger} et al.(2009a)]{MetzgerPQ09a} Metzger, B. D., Piro, A. L., \& Quataert, E. 2009a, MNRAS, 396, 304
\bibitem[{Metzger} et al.(2009b)]{MetzgerPQ09b} Metzger, B. D., Piro, A. L., \& Quataert, E. 2009b, MNRAS, 396, 1659
\bibitem[{Michel}(1988)]{Michel88} Michel, F. C. 1988, Nat, 333, 644
\bibitem[{Motl} et al.(2007)]{MotlFT07} Motl, P. M., Frank, J., Tohline, J. E., \& D'Souza, M. C. R. 2007, ApJ, 670, 1314
\bibitem[{Muno} et al.(2003)]{MunoBA03} Muno, M. P., Baganoff, F. K., \& Arabadjis, J. S. 2003, ApJ, 598, 474
\bibitem[{Muno} et al.(2005a)]{MunoLB05a} Muno, M. P., Lu, J. R., Baganoff, F. K., Brandt, W. N., Garmire, G. P., et al. 2005a, ApJ, 633, 228
\bibitem[{Muno} et al.(2005b)]{MunoPB05b} Muno, M. P., Pfahl, E., Baganoff, F. K., Brandt, W. N., Ghez, A., Lu, J., \& Morris, M. R. 2005b, ApJ, 622, L113
\bibitem[{Narayan} et al.(2001)]{NarayanPK01} Narayan, R., Piran, T., \& Kumar, P. 2001, ApJ, 557, 949
\bibitem[{Nomoto \& Iben}(1985)]{NomotoI85} Nomoto, K., \& Iben, I. Jr. 1985, ApJ, 297, 531
\bibitem[{Nomoto \& Kondo}(1991)]{NomotoK91} Nomoto, K., \& Kondo, Y. 1991, ApJ, 367, L19
\bibitem[{Ott}(2009)]{Ott09} Ott, C. D. 2009, Class. Quant. Gravity, 26, 063001
\bibitem[{Phinney}(1989)]{Phinney89} Phinney, E. S. 1989, in IAU Symp. 136, the Center of the Galaxy, ed. M. Morris (Dordrecht: Kluwer), 543
\bibitem[{Porquet}(2005)]{PGB05} Porquet, D., Grosso, N., Burwitz, V., et al. 2005, AA, 430, L9
\bibitem[{Pringle} et al.(1974)]{Pringle74} Pringle, J. E. 1974, Ph. D. thesis, University of Cambridge
\bibitem[{Pringle}(1981)]{Pringle81} Pringle, J. E. 1981, ARAA, 19, 137
\bibitem[{Pringle}(1991)]{Pringle91} Pringle, J. E. 1991, MNRAS, 248, 754
\bibitem[{Raskin \& Kasen}(2013)]{RaskinK13} Raskin, C., \& Kasen, D. 2013, ApJ, 772, 1
\bibitem[{Rees}(1988)]{Rees88} Rees, M. J. 1988, Nature, 333, 523
\bibitem[{Romanova} et al.(2004)]{RomanovaUK04} Romanova, M. M., Ustyugova, G. V., Koldoba, A. V., \& Lovelace, R. V. E. 2004, ApJ, 616, L151
\bibitem[{Romanova} et al.(2005)]{RomanovaUK05} Romanova, M. M., Ustyugova, G. V., Koldoba, A. V., \& Lovelace, R. V. E. 2005, ApJ, 635, L165
\bibitem[{Rueda} et al.(2013)]{RuedaBI13} Rueda, J. A., Boshkayev, K., Izzo, L., Ruffini, R., Lor$\acute{\rm e}$n-Aguilar, P., et al. 2013, ApJL, 772, L24
\bibitem[{Sadowski}(2011)]{Sadowski11} Sadowski, A. 2011, PhD thesis (Nicolaus Copernicus Astronomical Center); arXiv: 1108.0396
\bibitem[{Saio \& Jeffery}(2000)]{SaioJ00} Saio, H., \& Jeffery, C. S. 2000, MNRAS, 313, 671
\bibitem[{Saio \& Nomoto}(1985)]{SaioN85} Saio, H., \& Nomoto, K. 1985, AA, 150, L21
\bibitem[{Saio \& Nomoto}(1998)]{SaioN98} Saio, H., \& Nomoto, K. 1998, ApJ, 500, 388
\bibitem[{Saio \& Nomoto}(2004)]{SaioN04} Saio, H., \& Nomoto, K. 2004, ApJ, 615, 444
\bibitem[{Sakano} et al.(2005)]{SakanoWD05} Sakano, M., Warwick, R. S., Decourchelle, A., \& Wang, Q. D. 2005, MNRAS, 357, 1211
\bibitem[{Shakura \& Sunyaev}(1973)]{ShakuraS73} Shakura, N. I., \& Sunyaev, R. A. 1973, AA, 24, 337
\bibitem[{Shen \& Matzner}(2012)]{ShenM12} Shen, R.-F., \& Matzner, C. D. 2012, in EPJ Web of Conferences, Vol. 39, Tidal Disruption Events and AGN Outbursts, ed. R. Saxton \& S. Komossa (Madrid, Spain: EPJ Web of Conferences), 07006
\bibitem[{Toomre}(1964)]{Toomre64} Toomre, A. 1964, ApJ, 139, 1217
\bibitem[{Usov}(1992)]{Usov92} Usov, V. V. 1992, nature, 357, 472
\bibitem[{van Kerkwijk}(2013)]{vanKerk13} van Kerkwijk, M. H. 2013, Phil. Trans. R. Soc. A, 371, 20120236
\bibitem[{van Kerkwijk} et al.(2010)]{vanKerkCJ10} van Kerkwijk, M. H., Chang, P., \& Justham, S. 2010, ApJL, 722, L157
\bibitem[{Watarai} et al.(2000)]{WataraiFT00} Watarai, K., Fukue, J., \& Takeuchi, M. 2000, PASJ, 52, 133
\bibitem[{Watson} et al.(1985)]{WatsonKO85} Watson, M. G., King, A. R., \& Osborne, J. 1985, MNRAS, 212, 917
\bibitem[{Webbink}(1984)]{Webbink84} Webbink, R. F. 1984, ApJ, 277, 355
\bibitem[{Whelan \& Iben}(1973)]{WhelanI73} Whelan, J., \& Iben, I. 1973, ApJ, 186, 1007
\bibitem[{Wijnands} et al.(2006)]{WijnandsZR06} Wijnands, R., in't Zand, J. J. M., Rupen, M., Maccarone, T., Homan, J., Cornelisse, R., Fender, R., et al. 2006, AA, 449, 1117
\bibitem[{Wijnands} et al.(2002)]{WijnandsMW02} Wijnands, R., Miller, J. M., \& Wang, Q. D. 2002, ApJ, 579, 422
\bibitem[{Williams} et al.(2006)]{WilliamsNG06} Williams, B. F., Naik, S., Garcia, M. R., \& Callanan, P. J. 2006, ApJ, 643, 356
\bibitem[{Yoon} et al.(2007)]{YoonPod07} Yoon, S.-C., Podsiadlowski, P., \& Rosswog, S. 2007, MNRAS, 380, 933
\bibitem[{Yuan \& Narayan}(2004)]{YuanNara04} Yuan, F., \& Narayan, R. 2004, ApJ, 612, 724
\bibitem[{Yuan}(1992)]{Yuan92} Yuan, J. W. 1992, AA, 261, 105
\bibitem[{Yu \& Jeffery}(2010)]{YuJef10} Yu, S., \& Jeffery, C. S. 2010, AA, 521, A85
\bibitem[{Zhang} et al.(2014)]{ZhangJ14} Zhang, X., Jeffery, C. S., Chen, X., \& Han, Z. 2014, MNRAS, 445, 660
\bibitem[{Zhu} et al.(2013)]{ZhuCvan13} Zhu, C., Chang, P., van Kerkwijk, M. H., \& Wadsley, J. 2013, ApJ, 767, 164


\end{thebibliography}


\end{document}